\documentclass[twocolumn,prl,amsmath,amssymb,showpacs]{revtex4}
\usepackage{graphicx}
\usepackage{bm}
\usepackage{latexsym}
\usepackage{amstext}
\usepackage{amsxtra}
\usepackage{verbatim}
\usepackage[latin9]{inputenc}
\usepackage[T1]{fontenc}
\usepackage[english]{babel}
\usepackage{epstopdf}
\usepackage{aeguill}

\newcommand{\ER}{E_{\text{R}}}
\newcommand{\tn}{\textnormal}

\newcommand{\dlat}{d_\tn{lat}}

\newcommand{\be}{\begin{equation}}
\newcommand{\ee}{\end{equation}}

\newcommand{\rb}{{\bf r}}

\begin{document}
\title{Confinement-induced collapse of a dipolar Bose-Einstein condensate}
\author{J. Billy$^{1}$, E. A. L. Henn$^{1}$, S. M\"{u}ller$^{1}$, T. Maier$^{1}$, H. Kadau$^{1}$, A. Griesmaier$^{1}$, M. Jona-Lasinio$^{2}$, L. Santos$^{2}$ and T. Pfau$^{1}$}
 \affiliation{
 $^1$ 5. Physikalisches Institut, Universit\"{a}t Stuttgart, Pfaffenwaldring
 57, 70569 Stuttgart, Germany\\
 $^2$ Institut f\"{u}r Theoretische Physik, Leibniz Universit\"{a}t Hannover,
 30167 Hannover, Germany}
 \date{\today}

\begin{abstract}

We report on the observation of the confinement-induced collapse dynamics of a dipolar Bose-Einstein condensate (dBEC) in a one-dimensional optical lattice. We show that for a fixed interaction strength the collapse can be initiated in-trap by lowering the lattice depth below a critical value. 
Moreover, a stable dBEC in the lattice may become unstable during the time-of-flight dynamics upon release, due to the combined effect of 
the anisotropy of the dipolar interactions and inter-site coherence in the lattice.
\end{abstract}


 \maketitle

The stability of many-body systems, either classical or quantum, is typically determined by inter-particle interactions. 
Crucially, a change in the interaction energy balance may drive the system into instability, followed by a collapse, 
as spectacularly exemplified by stellar supernova explosions. Such interaction-induced instabilities have been observed as well
in degenerate quantum gases, where a change in inter-atomic interactions by means of Feshbach resonances
has been shown to induce the so-called Bose-Nova explosion~\cite{CollapseCornell,TheoryCollapseCornell}.

Recent experiments are exploring the physics of systems with significant dipole-dipole interactions~(DDI), including highly magnetic atoms 
as chromium~\cite{Griesmaier,Beaufils}, dysprosium~\cite{DyBEC} and erbium~\cite{ErBEC}, and ultra-cold polar molecules~\cite{Jila}. 
Even for the more complex interaction landscape provided by the DDI, the instability and collapse of a dBEC 
may be driven by a change in the interatomic interactions, the remarkable difference being that the anisotropy of the DDI 
leads to different collapse dynamics depending on the initial trap shape~\cite{Collapse,Collapse2}. 

In this work we show that the collapse of a dBEC can be induced as well by a change in the external trapping potential, 
while keeping the interaction strength constant. Such confinement-induced collapse, contrary to the interaction-induced case, relies on the geometry-dependent stability of a dBEC~\cite{Koch2008,PRA}. Thus, it is a general feature of dipolar systems, which cannot be observed in non-dipolar BECs, 
nor to our knowledge, in any other many-body system. 

We investigate the collapse dynamics of a dBEC with a fixed short-range interaction strength, trapped in a 1D optical lattice. 
Starting with an initially stable dBEC, we drive the system into instability by reducing the depth of the lattice potential 
below the stability threshold that we have recently mapped~\cite{PRA}. We show that this change in the external confinement induces 
an in-trap collapse, revealed by strong in-trap atom losses. Moreover, we show that a stable dBEC in the lattice may become unstable and collapse 
in time-of-flight (TOF) upon release. This TOF-induced collapse is a general feature of dipolar BECs in optical lattices, 
which, as we show, results from both the anisotropic DDI and the coherence of the BEC in the lattice. As TOF imaging is a major measurement technique, especially in lattice gases, the TOF-induced collapse demonstrated here is a key issue to be considered in experiments with 
polar lattice gases.

\begin{figure}[ht]
\centerline{\includegraphics[width=1\linewidth]{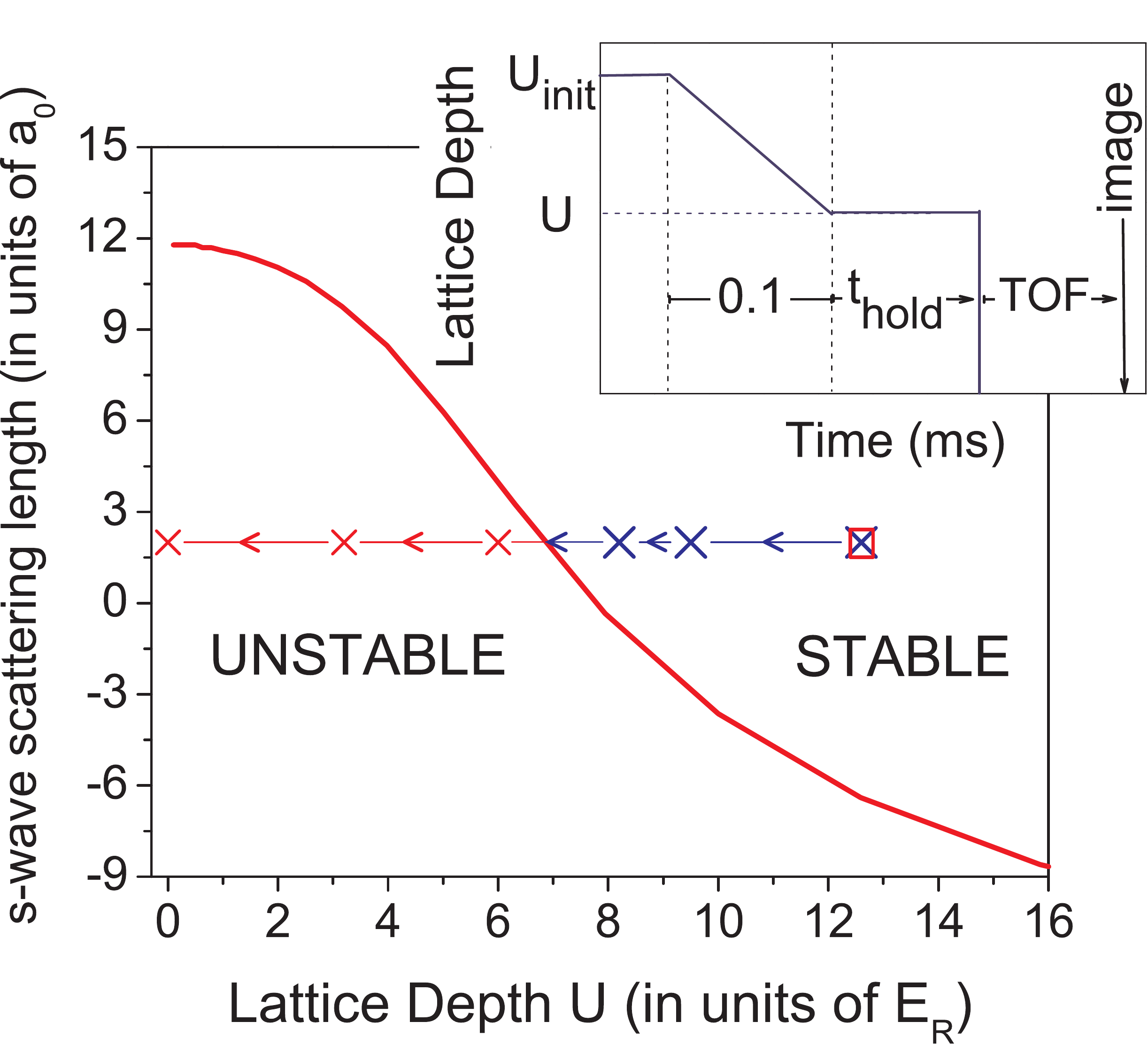}} 
\caption{Stability diagram of the $^{52}$Cr BEC trapped in the ODT and the 1D lattice. The red solid line corresponds to the stability threshold calculated for the experimental parameters~\cite{PRA}. The system is initially prepared in a stable configuration ($U_\mathrm{init} = 12.6\,\ER,a = 2 \pm 2\,a_\mathrm{0}$), denoted by the red square. The blue/dark gray (red/light gray) crosses correspond to the different values of the final lattice depth $U$ chosen in the stable (unstable) region. The inset depicts the time sequence of the experiment.}
\label{fig:TimelineExp}
\end{figure}

Our experimental procedure is as follows. We first prepare a quantum gas of bosonic $^{52}$Cr atoms in a stable configuration, following the sequence presented in Ref.~\cite{PRA}. The BEC, containing typically $15,000$ atoms, is confined in the combined potential produced by a crossed optical dipole trap (ODT), characterized by harmonic frequencies $\nu_{x,y,z}=\left(540,270,470\right)\tn{Hz}$, and a 1D optical lattice, with inter-site spacing $\dlat= 534\,\tn{nm}$, oriented along $z$. 
The atomic cloud is polarized by a strong magnetic field, also oriented along $z$, in the vicinity of a Feshbach resonance. 
We make use of this resonance to tune the $\textit{s-}$wave scattering length, characterizing the short-range contact interaction, down to $a = (2\pm 2)\, a_0$, with $a_0$ the Bohr radius. The lattice depth, initially equal to $U_\mathrm{init} =12.6\,\ER$ (where $\ER=\hbar^2\pi^2/(2m\dlat^2)$ 
is the recoil energy, with $m$ the atomic mass) is then ramped down to its final value $U$ in 100$\,\mu\tn{s}$, while keeping the scattering length constant. $U$ can be chosen arbitrarily above or under the stability threshold, the specific values that we are using in this paper being shown in Fig.~\ref{fig:TimelineExp}. We hold the system at this final configuration for an adjustable time $t_\mathrm{hold}$ and finally switch off all optical trapping potentials to perform an $8\,$ms TOF before taking an absorption image~\cite{Imaging}. 

\begin{figure*}[ht]
\centerline{\includegraphics[width=155mm]{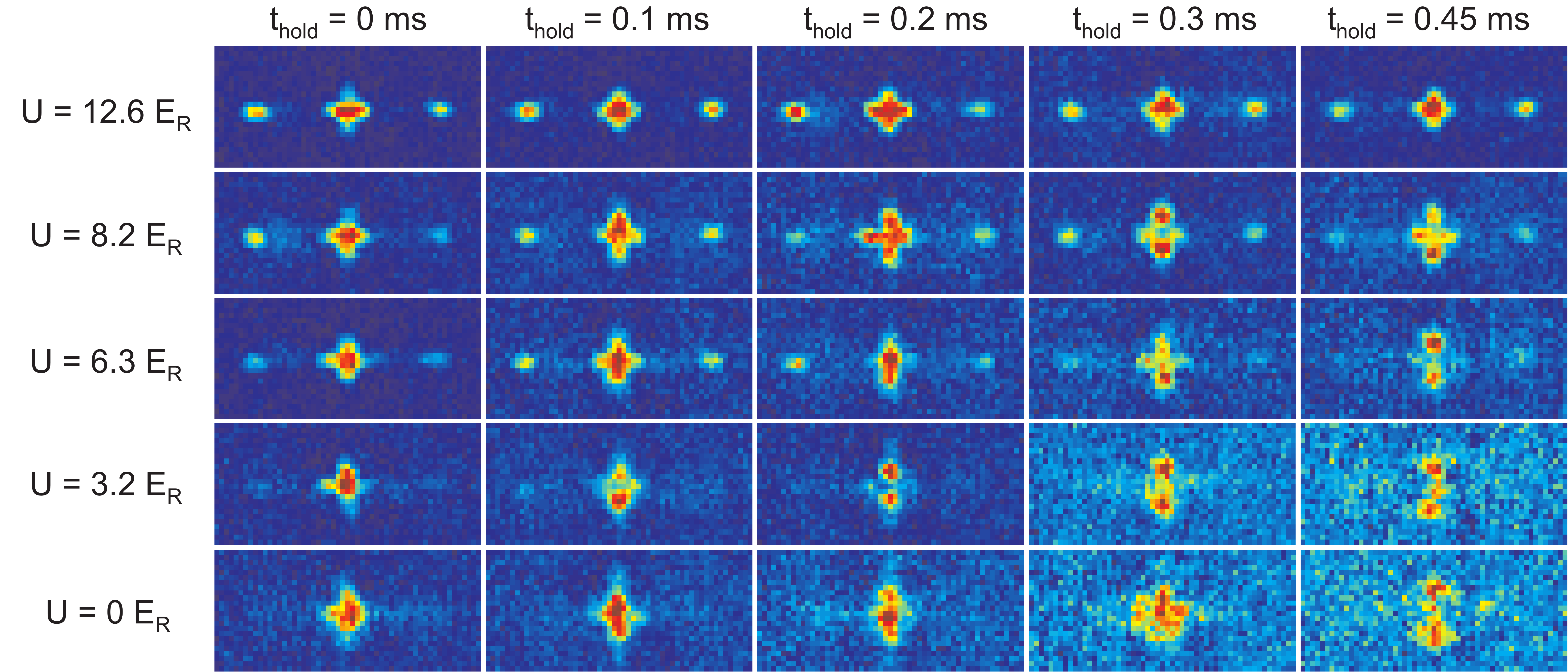}}
 \caption{Collapse dynamics: series of images (taken after an 8$\,$ms TOF) of the collapsing system for different in-trap holding times $t_\mathrm{hold}$, at different final lattice depth $U$ above and under the stability threshold (located around $7\,\ER$). Each image is obtained by averaging 5 absorption pictures, with the thermal cloud removed.}
\label{fig:ExpImages}
\end{figure*}

The time-evolution of the system is shown in Fig.~\ref{fig:ExpImages} for increasing holding time for different final lattice depths $U \leq U_\mathrm{init}$. Each snapshot results from the average of five absorption images taken after TOF under the same experimental conditions. Before averaging, the broad isotropic thermal cloud present on each single image was fitted by a Gaussian and subtracted from the image. Figure~\ref{fig:ExpImages} therefore shows the atomic patterns resulting from the interference of the remaining coherent atoms. 

\begin{figure}[h!]
\centerline{\includegraphics[width=1\linewidth]{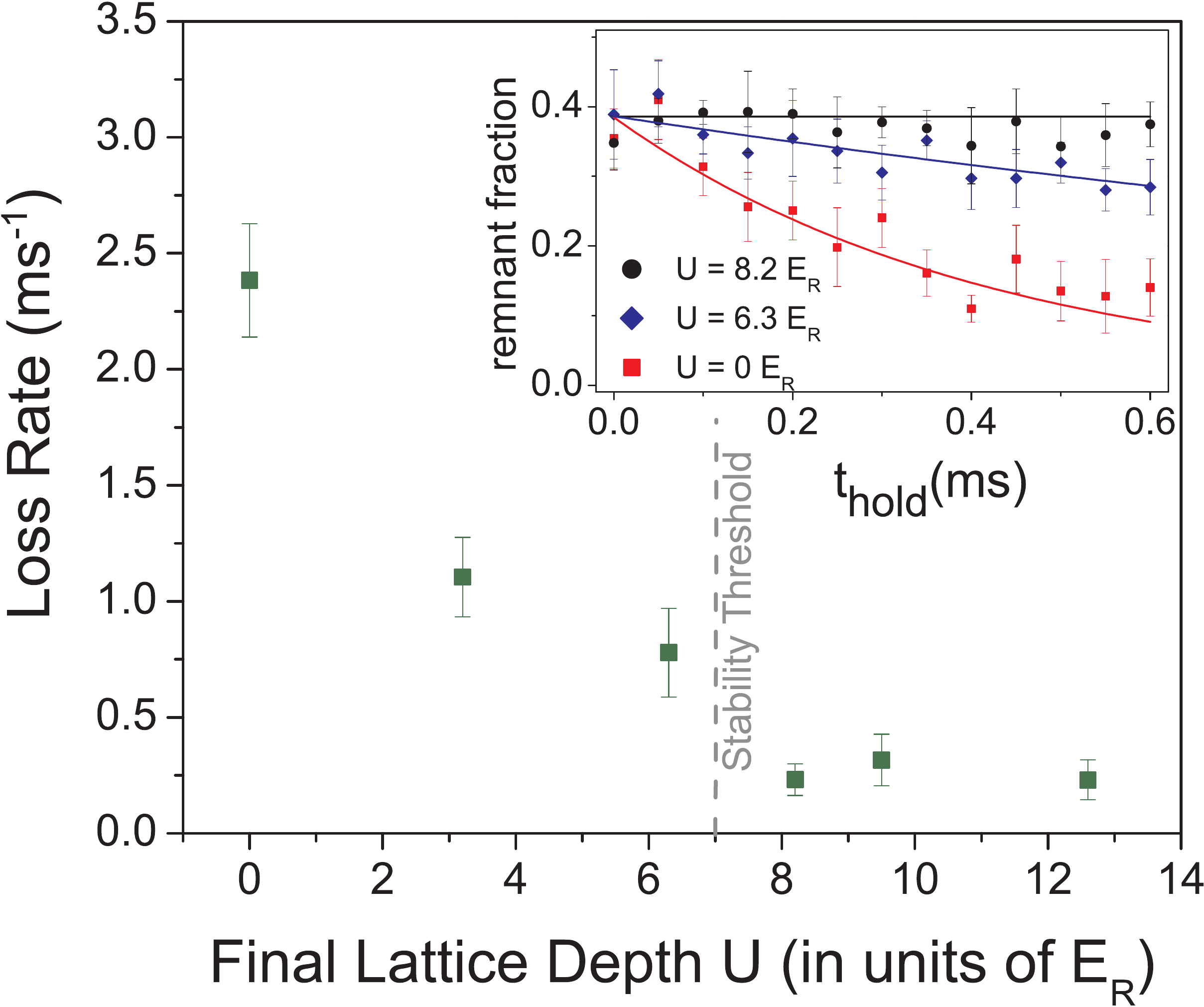}} 
\caption{Evolution of the loss rate of the remaining coherent atoms for different final lattice depths $U$. The loss rate strongly increases when crossing the stability threshold. The inset shows the time-evolution of the remnant fraction (see text for definition) for different final lattice depth $U$, from which the loss rates are extracted.}
\label{fig:AtomLosses}
\end{figure}

The system is either stable or unstable in-trap (i.e. before release) depending whether $U$ is chosen above or under the stability threshold, located at around 7$\,E_R$ for our parameters (see Fig.~\ref{fig:TimelineExp}). 
To distinguish between the different in-trap dynamics expected for these two regimes, 
we determine as a function of $t_\mathrm{hold}$ the evolution of the \textit{remnant fraction}, 
defined as the number of remaining coherent atoms (integrated over the images in Fig.~\ref{fig:ExpImages}) normalized to the total atom number before subtracting the thermal cloud~(see inset of Fig.~\ref{fig:AtomLosses}). 
From an exponential fit of the time evolution of the remnant fraction, we extract the atom loss rate, which serves as an observable for the in-trap dynamics. Loss rates are shown in Fig.~\ref{fig:AtomLosses} as a function of the final lattice depth $U$ before the TOF. 

We observe a clearly different dynamics depending on the value of $U$ compared to the stability threshold. For $U<7\,\ER$, the atomic cloud experiences an in-trap dynamics characterized by strong atom losses, showing that indeed the dBEC becomes unstable in-trap, before release. This collapse relies on the anisotropy of the DDI and the trap geometry after ramping down the lattice depth.

On the contrary, for $U>7\,E_R$ the system presents almost no evolution with the in-trap holding time $t_\mathrm{hold}$, 
as it can be seen from the very low loss rates. 
In addition, the sudden release from the lattice after the holding time results for all $t_\mathrm{hold}$ in the usual interference pattern formed by 
a central peak, corresponding to the zero-momentum component, and two side peaks, associated with the lattice recoil momentum 
$2\hbar k_\tn{lat}$ (with $k_\tn{lat} = \pi/\dlat$)~\cite{Morsch}. 
However, in contrast with typical interference patterns obtained from non-dipolar BECs, 
the central peak exhibits in our case a clear d-wave symmetry, similar to the one observed in interaction-induced 
collapse experiments~\cite{Collapse}. From the absence of evolution with $t_\mathrm{hold}$ and the observation of the d-wave shape of the central peak, we deduce that the collapse of the dBEC happens during the TOF: the system being stable before release, the collapse is therefore induced by the TOF itself~\cite{TOFcollapse}. As discussed below, 
this TOF-induced collapse is specific to dipolar gases in optical lattices and relies on the non-trivial interplay between the anisotropy 
of the dipolar interaction and the coherence of the system in the lattice. 

To examine more closely the condensate dynamics, we perform numerical simulations based on 
the nonlocal nonlinear Schr\"odinger equation:
\begin{multline} \label{eq:schrod1}
i\hbar \frac{\partial}{\partial t} \Psi(\rb,t) = \left[-\frac{\hbar^2}{2m} \nabla^2
+ V_{\rm ext}(\rb)
-i\hbar \frac{L_3}{2}N^2|\Psi(\rb,t)|^4\right. \\  \left.
+ N\int d\rb^\prime\,V_{\rm int}(\rb-\rb^\prime)|\Psi(\rb^\prime,t)|^2
\right]
\Psi(\rb,t)
\end{multline}
where $\Psi(\rb,t)$ is the condensate wavefunction, $m$ the atomic mass, and $N$ the 
initial number of atoms. The external potential 
$V_{\rm ext}(\rb)=  U \sin^2(\pi z/\dlat) + m \sum_{i=x,y,z}(2\pi\nu_i)^2 r_i^2/2$ 
results from the combination of the 1D optical lattice and the 3D harmonic confinement given by the ODT. The interaction energy is given by both contact and DDI potentials $V_{\rm int}(\rb) = \frac{4\pi\hbar^2 a}{m}\delta(\rb)
+\frac{\mu_0 \mu^2}{4\pi r^3} \left (1-3 \frac{z^2}{r^2} \right)
$, where the dipoles are polarized along $z$. The non-unitary term proportional
to the loss rate $L_3$ models non-resonant three-body losses, being essential for a realistic simulation of the collapse dynamics~\cite{Collapse,TheoryCollapseCornell}. $L_3$ is taken equal to $2\times 10^{-40}\,\text{m}^{6}/\text{s}$~\cite{Collapse}. The confinement-induced instability is then simulated by means of real-time evolution of Eq.~\eqref{eq:schrod1}, taking into account the whole experimental procedure described above, including the TOF expansion. 

\begin{figure*}[ht]
\centerline{\includegraphics[width=155mm]{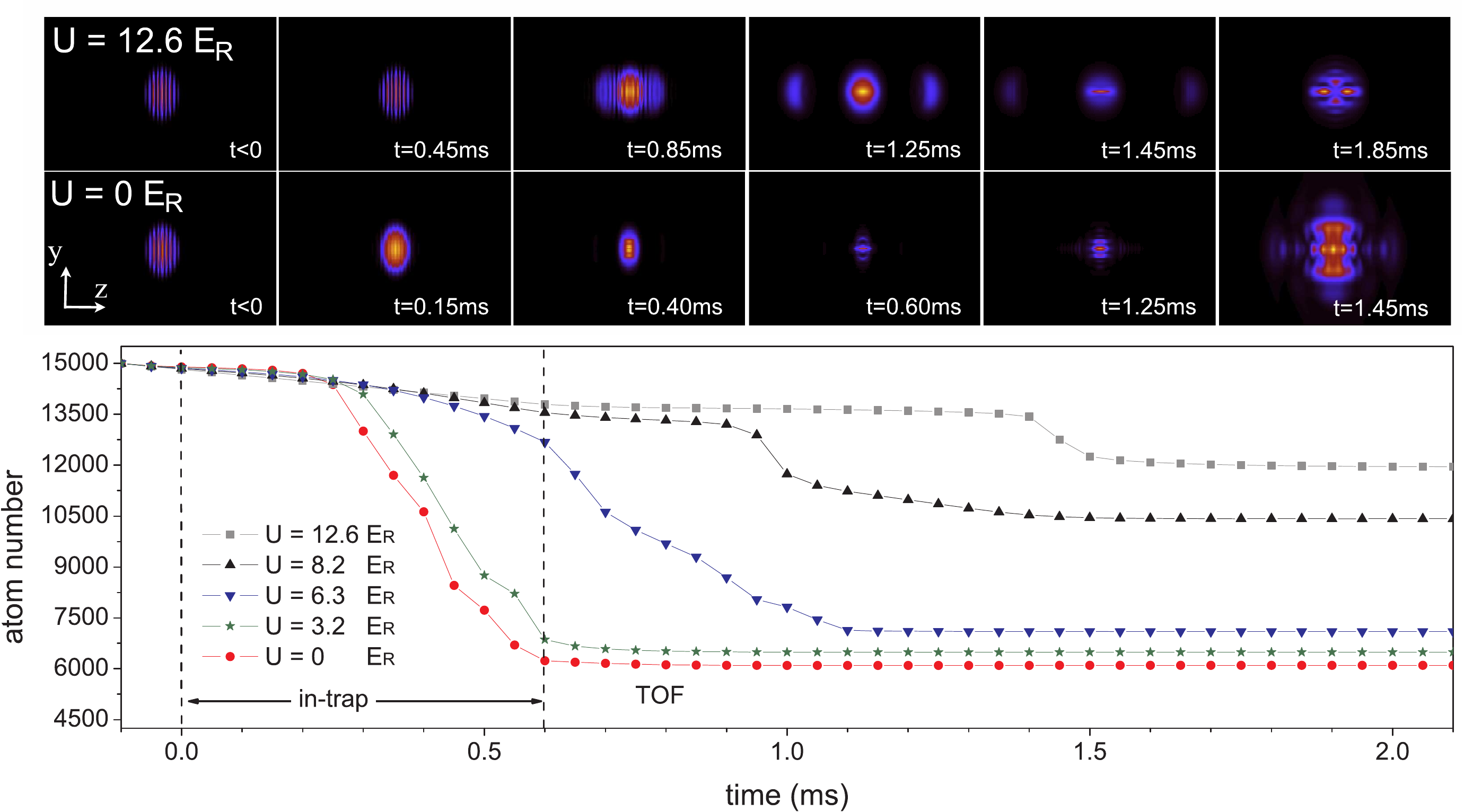}} 
\caption{Real-time simulations of the experimental collapse sequence shown Figure~\ref{fig:TimelineExp}. The first (second) row shows snapshots of the time-evolution of the system with a final lattice depth of $U = 12.6\,\ER$ ($U = 0\,\ER$). Note that the snapshots in the two rows are taken at different times. $t = 0$ marks the end of the lattice ramp and at t=0.6$\,$ms, the trapping potential is suddenly switched off. The lower panel wall shows the time-evolution of the atom number for different final lattice depths $U$, above as well as under the stability threshold.}
\label{fig:TofCollapse}
\end{figure*}

Results of our numerical simulations are presented in Figure~\ref{fig:TofCollapse}, where we show snapshots of the collapse dynamics of the system and the time evolution of the number of atoms, both for different final lattice depth $U$. Here we set the origin of the time axis at the end of the lattice ramp and we let the atomic cloud evolve for 0.6$\,$ms in-trap before releasing it. 

We first focus on the extreme case $U=0\,E_R$ below the stability threshold. In this case, we observe that the atomic cloud shrinks and undergoes strong atom losses while it is still trapped. Once released, however, the atomic cloud does not suffer any atom loss anymore. This shows indeed that the instability of the atomic cloud is initiated in-trap and that the subsequent in-trap collapse dynamics is associated with strong atom losses. After collapsing in-trap, the atomic cloud acquires a shape similar to the one observed experimentally and clearly visible after release. The dynamics of the system is different when $U$ gets closer to the stability threshold. This is particularly clear for $U=6.3\,\ER$: in this case, our simulations show that the atom losses start while the system is still trapped but only stops well after the beginning of the TOF. Therefore, even though the system starts to evolve in-trap, it only collapses with a d-wave symmetry during the TOF due to the short in-trap holding time. 

In contrast, in the case $U>7\,\ER$, we observe that the system is basically not evolving in-trap and undergoes only very low in-trap atom losses. This behavior reflects the stability of the atomic cloud before the TOF. In addition, our simulations show that the evolution of the system after release from the trap is a two-step process: first the high-momentum components of the wave function separate from its zero-momentum component. Then the latter shrinks transversally and collapses with a d-wave symmetry. This collapse of the central cloud can also be observed from the atom number, which exhibits a sudden and localized decrease during the TOF (occurring at around $t=1.45\,$ms for $U=12.6\,\ER$). Hence, our numerical simulations confirm the existence of the two types of collapse: an in-trap collapse and a TOF-induced collapse. 

We now consider the particular case of the TOF-induced collapse. To get more insight into this peculiar phenomenon, let us consider a deep enough lattice, 
such that the single-band approximation holds. We consider also for simplicity a decoupling of the radial~($x$, $y$) 
and axial~($z$) coordinates. Assuming inter-site coherence, the in-trap momentum distribution along the lattice direction is 
given by a series of narrow peaks at $k_l=2k_\tn{lat}l$ (with $l \in \mathbb{Z}$), whose form is given by $\widetilde F\left(k_z \right )$, 
the Fourier transform of the spatial envelope $F(z)$ associated to the overall harmonic confinement along $z$. 
Since the recoil energy is much larger than any other energy scale in the problem, at the early stages of the TOF~($t_\mathrm{tof}\sim \hbar/\ER$) particles belonging to different momentum peaks quickly move away from each other with a large relative velocity $2\hbar k_\text{lat}/m$. 
On the contrary the form of the wave packet associated to each individual peak evolves much more slowly, being given during these early TOF stages 
by the original envelope $F(z)$. The populations of the different wave packets are given by $N|\widetilde W_l|^2$, where $\widetilde W_l\equiv \widetilde W(k_l)$ is the Fourier transform of the on-site Wannier wave function. 

Each wave packet evolves then independently. Note that, crucially, interactions remain relevant for 
$t_\mathrm{tof}> \hbar/\ER$ due to the slow expansion dynamics of each individual wave packet. As a result, the wave packets may either expand indefinitely or collapse, depending on their population and geometry~(due to the anisotropy of the DDI). For our typical parameters, a variational analysis following a similar Gaussian ansatz as that introduced in Ref.~\cite{Yi2000} shows, in agreement with our experimental results, that the central peak collapses whereas the side peaks do not. This results from the different relative populations of the momentum peaks, which can be easily calculated by approximating the Wannier function by a Gaussian. For $U=12\,\ER$, the relative population of the zero-momentum component $|\widetilde W_0|^2=0.61$ is much higher than the relative population of each $2\hbar k_\tn{lat}$-momentum peak $|\widetilde W_1|^2=0.19$, leading to the collapse of the central peak only.

Note that the role of inter-site coherence is crucial in the above discussion. Indeed, in absence of coherence, the in-trap momentum distribution presents 
no individual peaks, but rather a broad Gaussian-like distribution $\widetilde W(k)$. As a result, after a time scale $\sim \hbar/\ER$, 
the incoherent sum of the expanded wave functions of each lattice site results in a single broad wave packet with a rapidly growing width. Thus, within this time scale, the atomic density drops dramatically, interactions become irrelevant and the atomic cloud expands freely. In addition, the collapse discussed above crucially depends on the anisotropy of the DDI, and the change of the cloud geometry in TOF. In particular, non-dipolar BECs, even with $a<0$, do not collapse in TOF if they were stable in-trap.

Finally, we note that confinement-induced collapse, either in-trap or TOF-induced, 
can be generalized to other dipolar systems as long as the scattering length of the system 
is chosen below $a_\mathrm{dd}$, where $a_\mathrm{dd}=m\mu_\mathrm{0}\mu^{2}/12\pi\hbar^{2}$ is the length scale 
associated to the DDI, with $\mu$ the magnetic dipole moment. In the case of chromium, $\mu=6\,\mu_B$ leading to 
$a_\mathrm{dd}\simeq 15\,a_\mathrm{0}$ and as expected, we do not observe any collapse arising when the 
same experiment is performed on a dBEC with a scattering length above $15\,a_\mathrm{0}$. 

In conclusion, we have shown that a dBEC may collapse under a change in its trapping potential. 
Such confinement-induced collapse, performed at constant interaction strength, 
is in strong contrast with previously studied interaction-induced collapses, being a characteristic feature of dBECs. 
We have furthermore shown that a dBEC initially stabilized in a 1D optical lattice can exhibit two different types of confinement-induced collapse, 
depending on the confinement configuration before release. On the one side, a dBEC may be destabilized while still trapped~(in-trap collapse). 
On the other side, a stable dBEC in the lattice may become unstable and collapse in TOF~(TOF-induced collapse). 

TOF-induced collapse is a characteristic feature of dBECs in lattices of moderated depths resulting from the anisotropy of the DDI and inter-site coherence. It shows that, contrary to the typical assumption that interactions do not play any role in TOF or only introduce a distortion in the momentum distribution, the TOF dynamics of a dBEC is more complex than expected. Since TOF imaging is a basic tool in ultra-cold gases, especially in lattice experiments, we stress that the TOF-induced collapse demonstrated here will have important consequences for future experiments on polar lattice gases, constituted either by atoms~\cite{DyBEC,ErBEC} or molecules~\cite{Jila}.

We thank K. Rz\k{a}\.{z}ewski for fruitful discussions. The Stuttgart group is supported by the German Research Foundation (DFG, through SFB/TRR21) and contract research `Internationale Spitzenforschung II' of the Baden-W\"{u}rttemberg Stiftung. Both groups acknowledge funding by the German-Israeli Foundation and the Cluster of Excellence QUEST. M.J.-L. and L.S. acknowledge financial support by the DFG (SA1031/6) and E.A.L.H. acknowledges support by the Alexander von Humboldt-Foundation.



\begin{thebibliography}{99}
\bibitem{Imaging}
The strong Feshbach magnetic field is switched-off after $4\,\tn{ms}$ of TOF and the atomic cloud is imaged by shining a resonant laser beam propagating along $x$ in the presence of a low magnetic field oriented along $x$~\cite{Collapse}.
\bibitem{TOFcollapse}
The TOF-collapse that we observe is fundamentally different from the interaction-induced collapse shown in Ref.~\cite{Collapse}, which happened partly during the TOF because the ramp in scattering length used to induce the collapse had not converged yet to its final value due to eddy currents.
\end{thebibliography}
\end{document}